%% file: main_arxiv.tex
\documentclass[twocolumn]{aastex63}
\usepackage{url}

\newcommand{\Pab}{Pa$\beta$}
\newcommand{\Pag}{Pa$\gamma$}
\newcommand{\Ha}{H$\alpha$}
\newcommand{\kms}{${\rm km\, s^{-1}}$}

\newcommand{\Vd}{$V_{\rm 1d}$}
\newcommand{\Kd}{$K_{\rm 1d}$}
\newcommand{\Vrot}{$V_{\rm rot}$}
\newcommand{\Mone}{M_1}
\newcommand{\Mtwo}{M_2}
\newcommand{\Kone}{K_1}
\newcommand{\Ktwo}{K_2}

\shorttitle{Disk emission lines in LB-1}
\shortauthors{Liu et al.}

\begin{document}

\title{Phase-dependent study of near-infrared disk emission lines in LB-1}

\author{Jifeng Liu}
\affiliation{Key Laboratory of Optical Astronomy,
National Astronomical Observatories, Chinese Academy of Sciences,
Beijing, China}
\affiliation{School of Astronomy and Space Sciences,
University of Chinese Academy of Sciences,
Beijing, China}
\affiliation{WHU-NAOC Joint Center for Astronomy,
Wuhan University,
Wuhan, China}

\author{Zheng Zheng}
\affiliation{Department of Physics and Astronomy, University of Utah, Salt Lake City, UT 84112, USA}

\author{Roberto Soria}
\affiliation{School of Astronomy and Space Sciences,
University of Chinese Academy of Sciences,
Beijing, China}
\affiliation{Sydney Institute for Astronomy,
The University of Sydney,
Sydney, NSW 2006, Australia}

\author{Jesus Aceituno}
\affiliation{Calar Alto Obsevatory, Spain}
\affiliation{Instituto de Astrofísica de Andalucia, Glorieta de la Astronomía, s/n, 18008, Granada, Spain}

\author{Haotong Zhang}
\affiliation{Key Laboratory of Optical Astronomy,
National Astronomical Observatories, Chinese Academy of Sciences, Beijing, China}

\author{Youjun Lu}
\affiliation{Key Laboratory of Optical Astronomy,
National Astronomical Observatories, Chinese Academy of Sciences, Beijing, China}
\affiliation{School of Astronomy and Space Sciences,
University of Chinese Academy of Sciences,
Beijing, China}


\author{Song Wang}
\affiliation{Key Laboratory of Optical Astronomy, National Astronomical Observatories, Chinese Academy of Sciences, Beijing, China}

\author{Wolf-Rainer Hamann}
\affiliation{Institut f\"ur Physik und Astronomie, Universit\"at Potsdam, Karl-Liebknecht-Str. 24/25, 14476, Potsdam, Germany}
\author{Lida M.~Oskinova}
\affiliation{Institut f\"ur Physik und Astronomie, Universit\"at Potsdam, Karl-Liebknecht-Str. 24/25, 14476, Potsdam, Germany}
\author{Varsha Ramachandran}
\affiliation{Institut f\"ur Physik und Astronomie, Universit\"at Potsdam, Karl-Liebknecht-Str. 24/25, 14476, Potsdam, Germany}

\author{Hailong Yuan}
\affiliation{Key Laboratory of Optical Astronomy,  National Astronomical Observatories, Chinese Academy of Sciences, Beijing, China}
\author{Zhongrui Bai}
\affiliation{Key Laboratory of Optical Astronomy, National Astronomical Observatories, Chinese Academy of Sciences, Beijing, China}
\author{Shu Wang}
\affiliation{Key Laboratory of Optical Astronomy, National Astronomical Observatories, Chinese Academy of Sciences, Beijing, China}
\author{Brendan J.~McKee}
\affiliation{Sydney Institute for Astronomy, 
The University of Sydney, 
Sydney, NSW 2006, Australia}
\author{Jianfeng Wu}
\affiliation{Department of Astronomy, Xiamen University, Xiamen, China}
\author{Junfeng Wang}
\affiliation{Department of Astronomy, Xiamen University, Xiamen, China}
\author{Mario Lattanzi}
\affiliation{INAF--Osservatorio Astrofisico di Torino, Strada Osservatorio 20, 10025 Pino Torinese TO, Italy}
\author{Krzysztof Belczynski}
\affiliation{Nicolaus Copernicus Astronomical Centre, Polish Academy of Sciences, ul. Bartycka 18, PL-00-716 Warsaw, Poland}
\author{Jorge Casares}
\affiliation{Instituto de Astrof\'{i}sica de Canarias, c/V\'{i}a L\'{a}ctea s/n, E-38200 La Laguna , Tenerife, Spain}
\affiliation{Departamento de Astrofísica, Universidad de La Laguna, E-38206, Santa Cruz de Tenerife, Spain}
\author{Sergio Simon-Diaz}
\affiliation{Instituto de Astrof\'{i}sica de Canarias, c/V\'{i}a L\'{a}ctea s/n, E-38200 La Laguna , Tenerife, Spain}
\author{Jonay I. Gonz\'{a}lez Hern\'{a}ndez}
\affiliation{Instituto de Astrof\'{i}sica de Canarias, c/V\'{i}a L\'{a}ctea s/n, E-38200 La Laguna , Tenerife, Spain}
\author{Rafael Rebolo}
\affiliation{Instituto de Astrof\'{i}sica de Canarias, c/V\'{i}a L\'{a}ctea s/n, E-38200 La Laguna , Tenerife, Spain}




\begin{abstract}
The mass, origin and evolutionary stage of the binary system LB-1 has been the subject of intense debate, following the claim that it hosts an $\sim$70$M_{\odot}$ black hole, in stark contrast with the expectations for stellar remnants in the Milky Way. 
We conducted a high-resolution, phase-resolved spectroscopic study of the near-infrared Paschen lines in this system, using the 3.5-m telescope at Calar Alto Observatory. We find that \Pab\ and \Pag\ (after proper subtraction of the stellar absorption component) are well fitted with a standard double-peaked model, typical of disk emission. We measured the velocity shifts of the red and blue peaks at 28 orbital phases: the line center has an orbital motion in perfect antiphase with the stellar motion, and the radial velocity amplitude ranges from 8 to 13 km/s for different choices of lines and profile modelling. 
We interpret this curve as proof that the disk is tracing the orbital motion of the primary, ruling out the circumbinary disk and the hierarchical triple scenarios. The phase-averaged peak-to-peak half-separation (proxy for the projected rotational velocity of the outer disk) is $\sim$70 km s$^{-1}$, larger than the stellar orbital velocity and also inconsistent with a circumbinary disk. 
From those results, we infer a primary mass 4--8 times higher than the secondary mass. Moreover, we show that the ratio of the blue and red peaks (V/R intensity ratio) has a sinusoidal behaviour in phase with the secondary star, which can be interpreted as the effect of external irradiation by the secondary star on the outer disk.
Finally, we briefly discuss our findings in the context of alternative scenarios recently proposed for LB-1. Definitive tests between alternative solutions will require further astrometric data from {\it Gaia}.  

\end{abstract}

\keywords{stars: black holes; stars: evolution; stars: emission-line, Be}


\section{Introduction}
\label{sec:intro}

The traditional way to discover and identify stellar-mass black holes (BHs) in binary systems is to spot those that are undergoing significant gas accretion, a process associated with emission of high-energy photons; the X-ray discovery usually comes first, followed by studies of their optical and multiband counterparts (more suitable for the measurement of binary period, mass functions, and other system parameters). The limitation of this approach is that only a small fraction of BHs is X-ray bright. So far, only about two dozen X-ray active BHs have been identified in the Milky Way \citep{corral-santana16,kreidberg12}, with another 40 possible candidates\footnote{\url{http://www.astro.puc.cl/BlackCAT/}}. This small sample may not be statistically representative of the whole population of $\sim$10$^8$ such objects floating around in our galaxy. If so, their empirical properties may erroneously skew theoretical efforts to model for example massive stellar evolution and collapse.

To address this problem, in 2016 we started a search for spectroscopic binaries with the Large Sky Area Multi-Object Fiber Spectroscopic Telescope (LAMOST), at Xinglong Observatory. Our goal was to find BHs in stellar-mass binaries directly from optical radial velocity measurements. We monitored about 3000 targets in the {\it Kepler} K2-0 field. Among this sample, \cite{liu19} (henceforth, L19) discovered the very interesting binary system LB-1 (LS V $+22$ 25 in the {\sc{simbad}} database), with a {\it Gaia} J2000 position of R.A. = 06$^{\rm h}$11$^{\rm m}$49$^{\rm s}$.076 and Dec. = $+$22$^{\circ}$49$^{\prime}32\farcs66$ \citep{gaia18}, corresponding to Galactic coordinates $(l,b)=(188^\circ.23526,2^\circ.05089)$.
After the LAMOST discovery, L19 monitored the system between 2017 December and 2018 April also with the OSIRIS spectrograph on the Gran Telescopio Canarias (GTC), and with the HIRES spectrograph on Keck I. 

Based on the modelling of those spectra, L19 claimed the following main results: i) the stellar absorption lines have a perfectly sinusoidal velocity curve with a period $P_{\rm {bin}} = 78.9 \pm 0.2$ d and semi-amplitude $\Ktwo = 52.8 \pm 0.7$ km s$^{-1}$;  ii) the absorption line spectrum is consistent\footnote{Based on the \sc{tlusty} model, \cite{hubeny95}.} with a B3 star with a temperature $T_{\rm {eff}} = 18,100 \pm 820$ K, surface gravity 
$\log[g/{\rm (cm\  s^{-2})}] = 3.43 \pm 0.15$ and mass $\Mtwo = (8.2 \pm 0.9) M_{\odot}$; iii) the broad width and the wine-bottle shape of the H$\alpha$ emission line suggest its origin from an accretion disk around a BH viewed nearly pole-on; iv) the H$\alpha$ emission line wings show a BH motion in anti-phase with the B star itself, with a semi-amplitude $\Kone = 6.4 \pm 0.8$ km s$^{-1}$; v) from the ratio of velocity amplitude, the BH mass is $\Mone \equiv \left(\Ktwo/\Kone\right)\, \Mtwo = 68^{+11}_{-13} M_{\odot}$.

A mass $>$55 $M_{\odot}$ is at odds with the most commonly expected upper limit of BH masses in a solar-metallicity environment ($M \lesssim 25 M_{\odot}$ in \citealt{spera15}; $M \lesssim 35 M_{\odot}$ in \citealt{spera17} ). Moreover, the existence of such a BH would contradict or at least push the boundaries of the theoretical prediction of a mass gap, where a collapsing stellar core is disrupted in a pair-instability pulsation supernova or a pair instability supernova, without the formation of a BH \citep{heger03,spera17,woosley17,farmer19,leung19}. On the other hand, it was proposed \citep{petit17,belczynski20} that single stars can produce a 70-$M_{\odot}$ BH even at solar metallicity, if the mass loss rate in the wind (prior to core collapse) is strongly reduced, for example because of surface magnetic fields. However, this scenario does not help in the case of LB-1, because the binary separation is not large enough to fit the stellar progenitor of such a massive BH  \citep{belczynski20}. Thus, we must consider the possibility that the BH mass claimed by L19 was over-estimated or even that it does not contain a BH at all.

\begin{deluxetable}{ccccc}
\tablenum{1}
\tablecaption{Log of our Calar Alto observations\label{tab:log}}
\tablehead{
\colhead{Obs Date} & \colhead{BJD$_{mid}$} & \colhead{Exp Time (s)} & \colhead{$\phi_{mid}$} & \colhead{Band}
}
{\footnotesize{
\startdata
2019-10-01  &  58757.6747  &  471  &  0.8828  &  NIR\\ 
            &  58757.6763  &  483  &  0.8829  &  VIS\\ 
            &  58757.6812  &  398  &  0.8829  &  NIR\\
            &  58757.6828  &  411  &  0.8830  &  VIS\\
2019-11-20  &  58807.6374  &  1797  &  0.5161  &  NIR\\    
            &  58807.6377  &  1800  &  0.5161  &  VIS\\
2019-11-24  &  58811.5836  &  1198  &  0.5661  &  NIR\\
            &  58811.5840  &  1242  &  0.5661  &  VIS\\
2019-11-28  &  58815.5388  &  396  &  0.6162  &  NIR\\
            &  58815.5389  &   404  &  0.6162  &  VIS\\
\enddata
}}
\tablecomments{
Phase $\phi = 0$ corresponds to inferior conjunction of the companion star, orbiting in a counterclockwise direction. Thus, $\phi = 0.25$ corresponds to its highest projected radial velocity.\\
(This table is available in its entirety in a machine-readable form in the online journal. A portion is shown here for guidance regarding
its form and content.)
}
\end{deluxetable}

Several authors have questioned the two main tenets of the L19 scenario: the modelling and interpretation of the disk emission lines \citep{Abdul2019,elbadry20}, and the mass of the star responsible for the narrow absorption lines \citep{Abdul2019,Simon2020,Irrgang2020,shenar20}. Both \cite{Abdul2019} and \cite{elbadry20} independently argued that the apparent sinusoidal motion of the H$\alpha$ emission line disappears when the H$\alpha$ absorption line from the companion star is properly subtracted, suggesting that the line comes from a circumbinary disk rather than from a disk inside the Roche lobe of the primary. 
Both \cite{Abdul2019} and \cite{Simon2020} estimated a B star mass of 3--6$M_\odot$, while \citet{Irrgang2020} suggested a stripped helium star of $\sim1M_\odot$ for the companion.
The possibility that the primary in LB-1 is a Be star rather than a BH was recently proposed by \cite{shenar20}, by analogy with other binary systems composed of a Be star plus a hot subdwarf, with similar binary periods of several months and similar line spectra (e.g., $\varphi$ Per: \citealt{gies98}; $o$ Pup: \citealt{koubsky12}). \cite{rivinius20} recently proposed that LB-1 is likely a triple star system analogous to HR 6819, which has a B3III+BH inner binary and a distant Be star giving apparently non-moving H$\alpha$ emission lines. 

In this paper, we will present new data and modelling, which may clarify the origin and location of the line emission. We will leave a direct investigation of the spectral type of the companion star to further work. Our main immediate objective for this work is to test whether the emission lines have a sinusoidal motion. To do so, we side-step all the intrinsic complications of the H$\alpha$ line profile (likely the result of multiple emission and absorption components, and modified by scattering), and examine instead the Paschen lines, in particular \Pab\ and \Pag. The Paschen lines have cleaner, simpler double-peaked, disk-like profiles. We do a phase-resolved study of such profiles based on new high-resolution spectroscopic data, and model them to determine the rotational velocity of the outer disk, the orbital velocity of the primary, and the mass ratio. We then briefly discuss which ones of the many proposed scenarios for LB-1 are most consistent with our new findings or not. 

Throughout the paper, we use subscripts `1' and  `2' to denote quantities related to the primary (unseen companion) and the secondary (star), respectively, such as mass $\Mone$ and $\Mtwo$, and orbital velocity semi-amplitude $\Kone$ and $\Ktwo$.

\section{New observations and data analysis} \label{sec:observations}

We have observed LB-1 at regular intervals (28 epochs) since November 2019 (Table~\ref{tab:log}), with the ``Calar Alto high-Resolution search for M dwarfs with Exoearths with Near-infrared and optical \'Echelle Spectrographs'' (CARMENES) mounted on the 3.5-m telescope at the Calar Alto Observatory.
The CARMENES instrument \citep{quirrenbach14,seifert16} was developed by a consortium of Spanish and German institutions. It consists of two separate spectrographs covering the wavelength ranges from 0.52 to 0.96 $\mu$m and from 0.96 to 1.71 $\mu$m, with spectral resolution $R \approx 80,000$--100,000. 
In this work, we make use of the near-IR spectral coverage to reveal the Paschen emission lines of LB-1.

Data reduction was performed automatically with the CARMENES Reduction And
CALibration ({\sc {caracal}}) pipeline software \citep{zechmeister14,caballero16}; this package is based on the {\sc{reduce}} software \citep{piskunov02} with additional scripts added or modified. Specifically, {\sc {caracal}} carries out dark and bias corrections, order tracing, flat-relative optimal extraction \citep{zechmeister14} and wavelength calibration \citep{bauer15}.
It produces wavelength-calibrated one dimension spectra. It also computes an approximate radial velocity by comparison with synthetic spectral templates of main-sequence stars. After correcting for the instrumental response, the pipeline software merges all the \'echelle orders in a single matrix, with a final output of wavelength, flux, error in flux and background values.

\begin{figure}
   \gridline{
        \fig{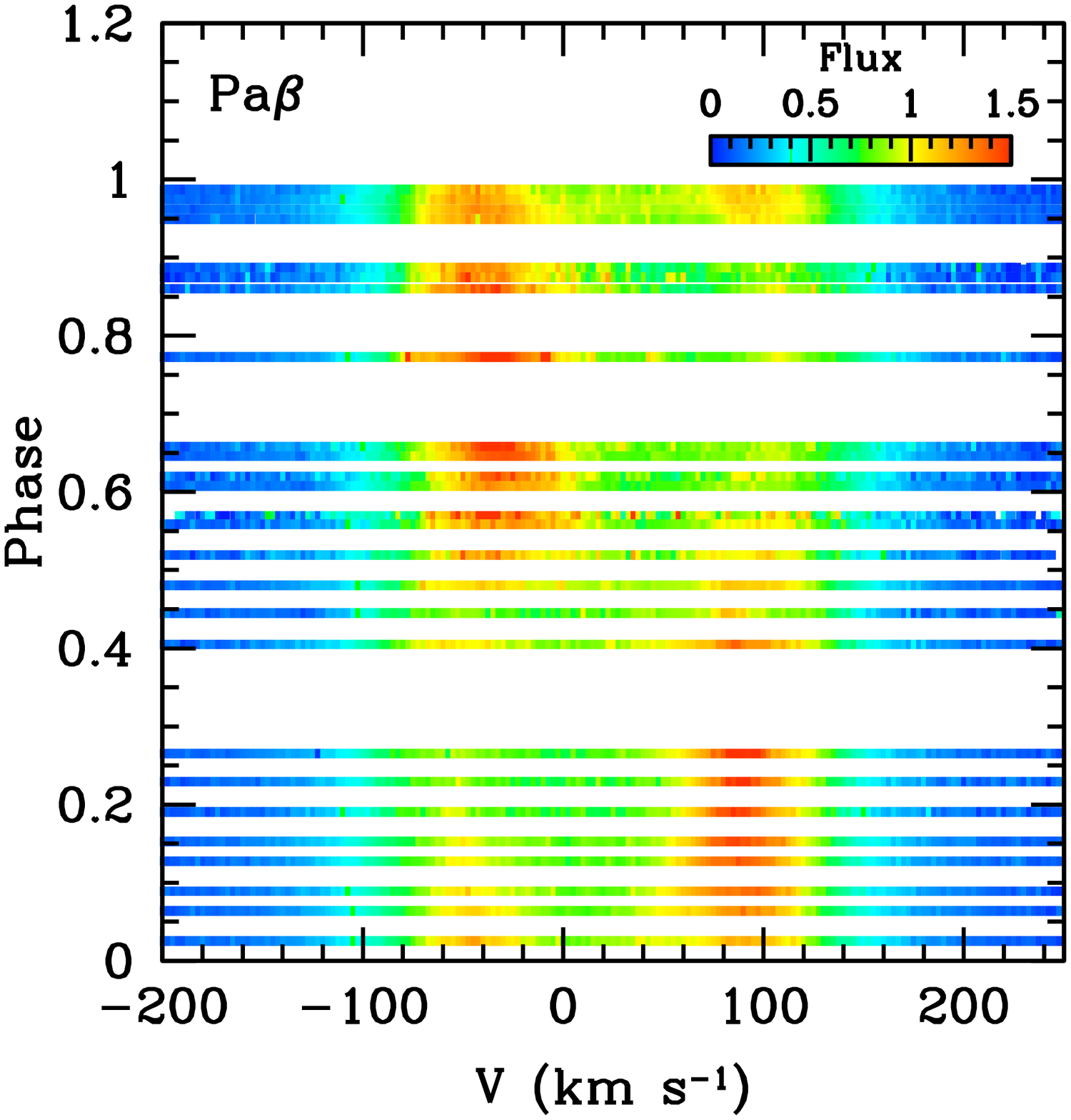}{0.48\linewidth}{}
        \fig{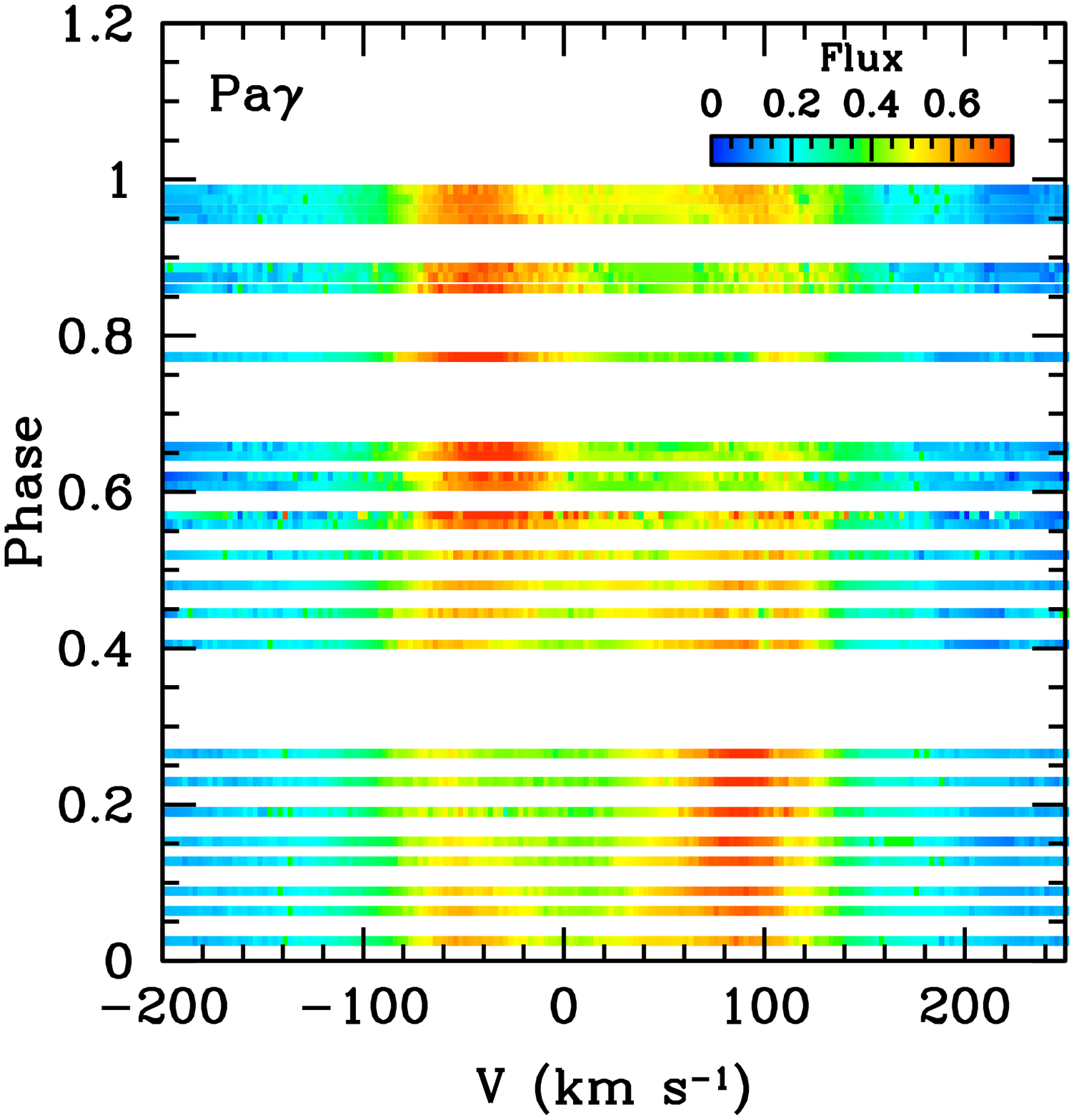}{0.48\linewidth}{}
    }
    \vspace{-0.8cm}
   \gridline{
        \fig{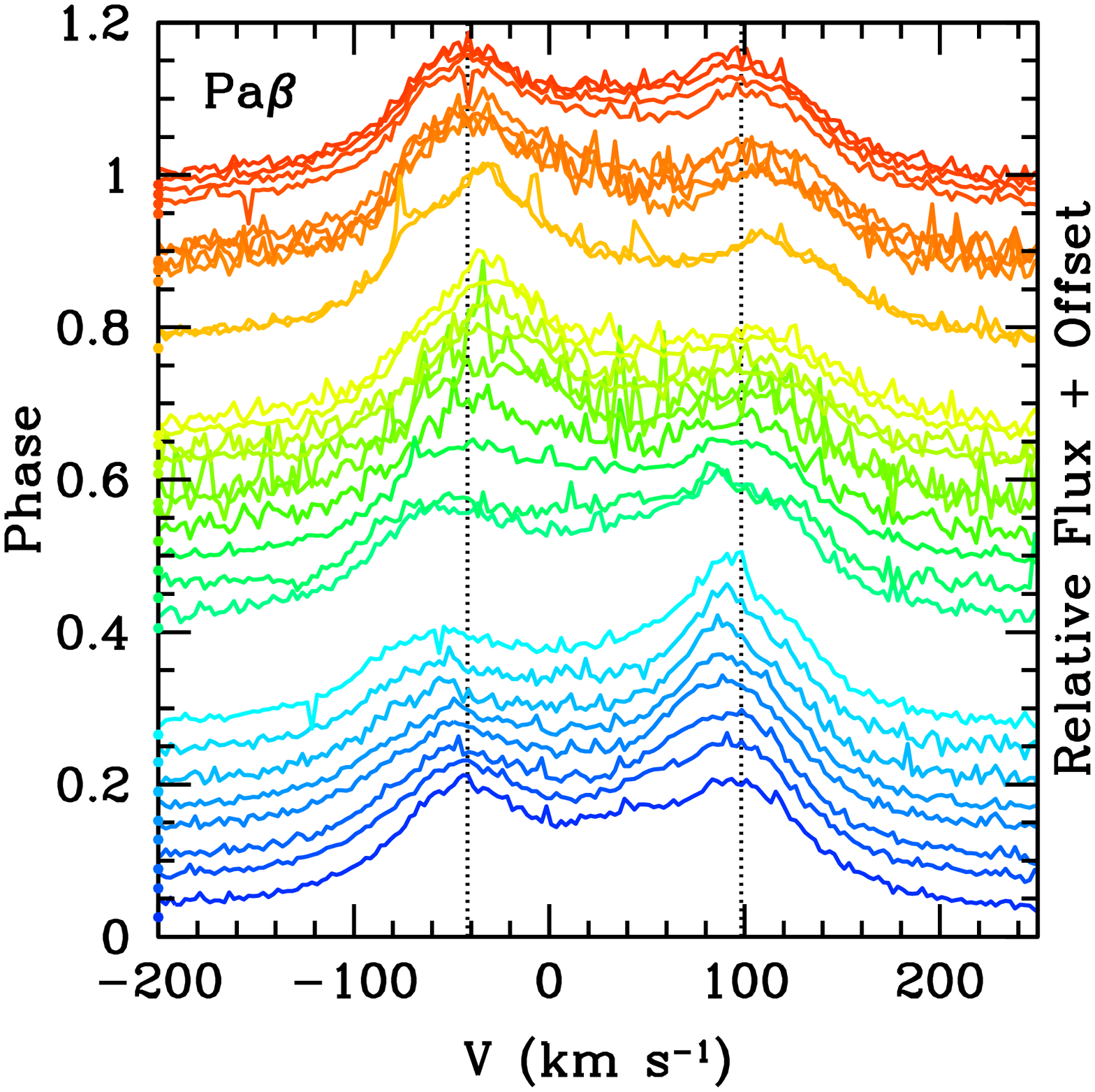}{0.48\linewidth}{}
        \fig{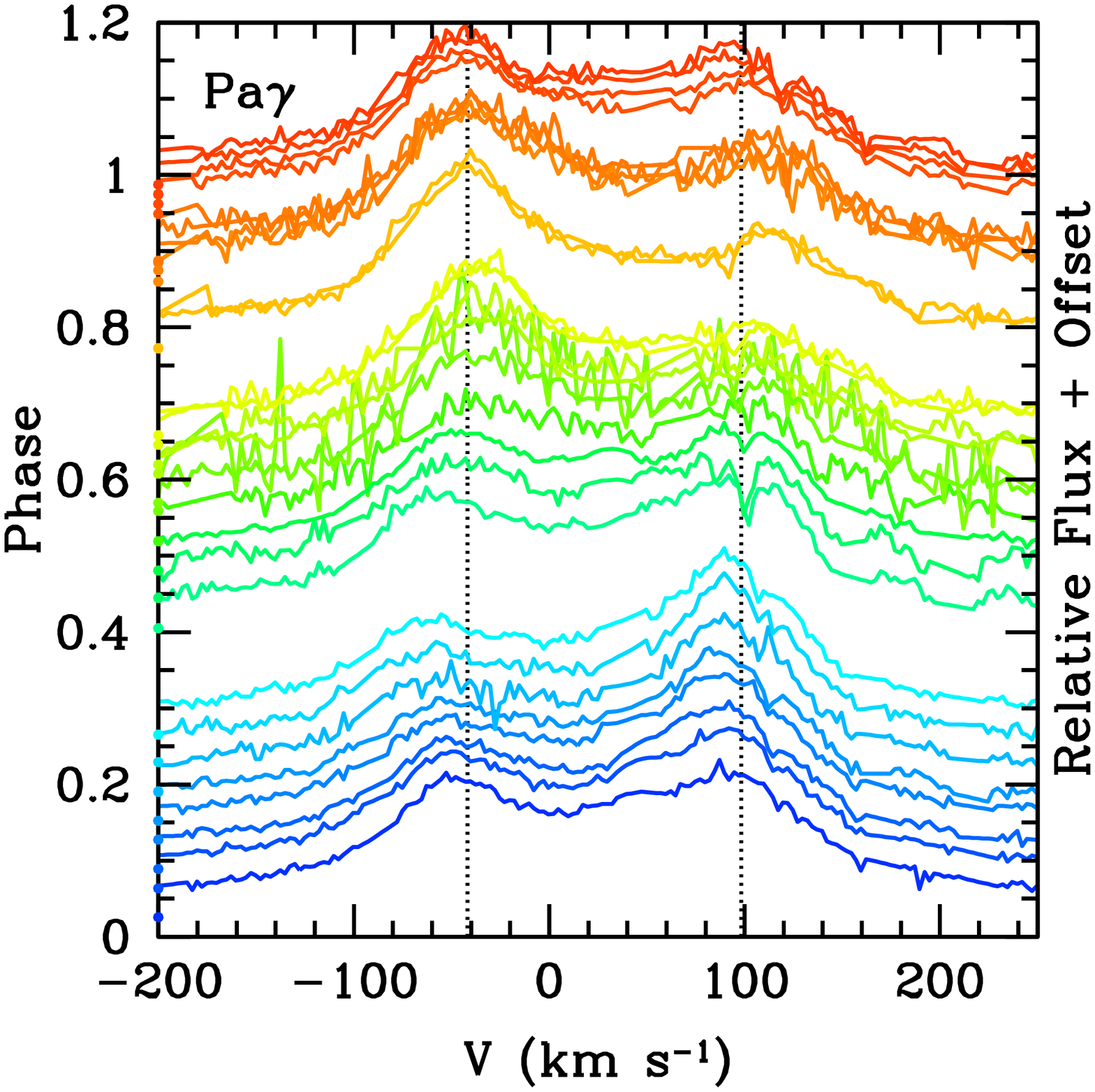}{0.48\linewidth}{}
    }
    \vspace{-0.5cm}
\caption{
Observed \Pab\ (left panels) and \Pag\ (right panels) emission line profiles, after subtraction of the phase-dependent stellar absorption components. For each line, the top panel shows the phase sequence of 28 emission line spectra from our CARMENES observations, plotted as a color map with flux in units of the stellar continuum near the line. The bottom panel plots the same sequence of line profiles as 1-d spectra, with all fluxes scaled by the same arbitrary factor for plotting purposes. We shifted the normalization baseline of each profile to the value of the corresponding binary phase, which is indicated by reference bullet points on the left vertical axis
(the reference points have the same color as the corresponding profiles). 
The dotted vertical lines are at $\pm$70 \kms\ from the center-of-mass velocity $V_{\rm sys}=28.3$ \kms\ of the binary system: they are plotted as visual guides to highlight the shifts in the line profiles. 
}
\label{fig:Pab}
\vspace{0.3cm}
\end{figure}

\section{Main Results}

\subsection{Radial velocity of the secondary star}

First, we determined the radial velocity amplitude of the secondary star, by fitting the position of the \ion{He}{1} absorption lines. We obtain a projected velocity semi-amplitude $\Ktwo = 52.62 \pm 0.15$ km s$^{-1}$, confirming the results of L19 but with higher precision. The eccentricity is consistent with 0 ($e \lesssim 0.01$). The systemic velocity (center-of-mass velocity of the binary system) is $V_{\rm sys} = 28.3 \pm 0.1$ km s$^{-1}$.

The ephemeris of the system is
\begin{equation}
T(\phi=0) = 58845.8(2) BJD + 78.9(0) \times N,
\end{equation}
where $\phi = 0$ corresponds to the secondary star in inferior conjunction with the primary object, and BJD is the Barycentric Julian Date.

An accurate measurement of the stellar motion enables us to subtract the stellar absorption component (properly shifted by the radial velocity offset corresponding to each phase) from the Paschen emission lines. For the absorption line profile, we use the Kurucz atmosphere model \citep{kurucz79} with $T_{\rm eff}=13,500$ K and $\log[g/{\rm (cm\  s^{-2})}]=3.5$. As we focus on analyzing the peaks of emission lines, our results are robust to different choices of absorption line profiles corresponding to a plausible range of temperatures (13,000--18,000 K). In particular, we have performed tests using different absorption profiles, calculated in two ways: i) with a Kurucz atmosphere model with $T_{\rm eff}=18,000$ K and $\log g=3.5$; ii) with the Potsdam Wolf-Rayet (PoWR) models \citep{sander15,hamann03,grafener02} with $T_{\rm eff}=13,500$ K, $\log g=2.8$, and extremely high mass-loss rate $\dot{M}=10^{-7.5}M_\odot {\rm yr}^{-1}$. The peak positions of the emission components and thus our main results are found to be essentially unaffected.

\begin{figure}
\includegraphics[width=\linewidth]{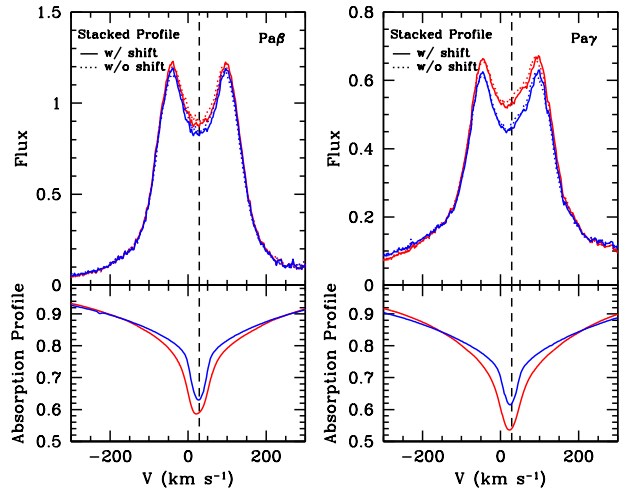}
\caption{
Stacked \Pab\ (left) and \Pag\ (right) emission line profiles.
Two different stellar absorption profiles are adopted to obtain the emission line profile, as shown in the bottom panels. The blue curves are the default ones, while the red curves represent an extreme case from a model with strong stellar wind (see the text for detail), both centered at the systemic velocity $V_{\rm sys}=28.3$ \kms (Section~\ref{sec:lineprof}), which is the center-of-mass velocity of the binary system inferred from the radial velocity curve of the B star. The dashed vertical line marks this systemic velocity.
For a given stellar absorption profile, each top panel shows two (almost identical) versions of a stacked line profile (after subtraction of the phase-dependent absorption profile). The dotted curve is the direct stack of the 28 observed profiles, weighted by inverse variance; the solid curve is a stack of the same profiles, but with a shift for each profile so that the average velocity of the two peaks is put at $V_{\rm sys}$. The flux is in units of the B-star continuum near the line.
}
\label{fig:stacked}
\vspace{0.3cm}
\end{figure}

\subsection{Profile and width of the Paschen emission lines} \label{sec:profiles}
\label{sec:lineprof}

\begin{figure*}
\includegraphics[width=\linewidth]{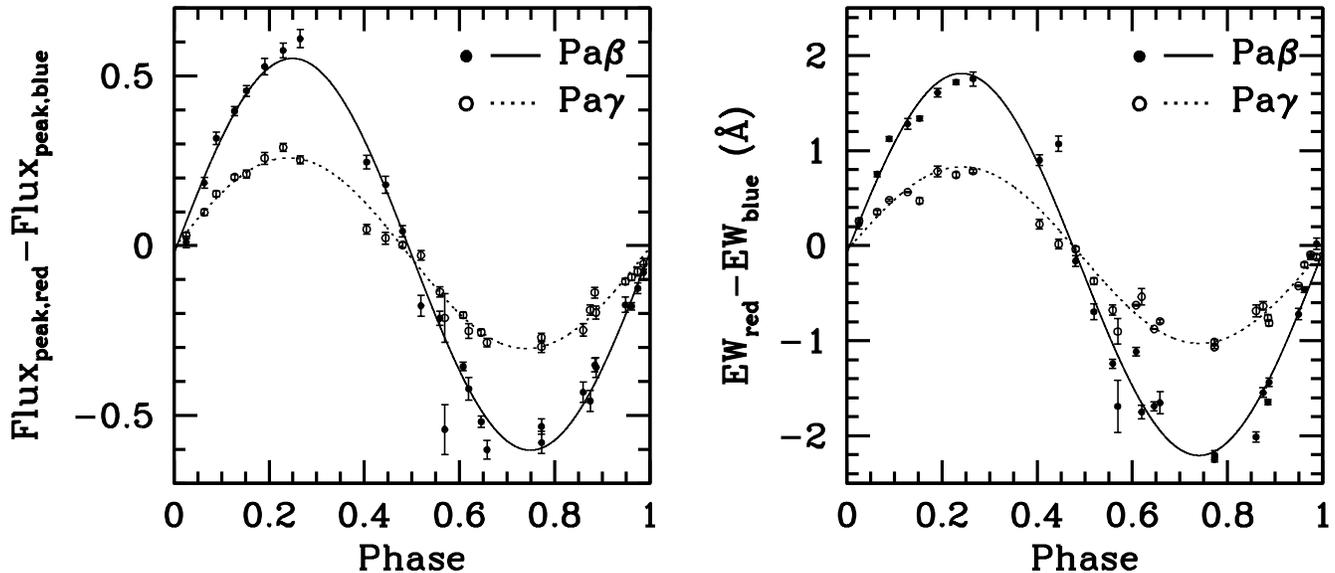}
\caption{
Asymmetry between the red- and blue-side emission as a function of binary phase, for both \Pab\ and \Pag. The left panel shows the difference between the red and blue peak fluxes; the right panel shows the difference between the equivalent widths redward and blueward of the mean peak locations. The datapoints are fitted with sinusoidal curves; the phases of those curves were left as free parameters in the fits, but they turn out to be consistent with 0 (i.e., with the phase of the secondary star). More exactly, the initial phases for \Pab\ are $0.001 \pm 0.003$ (left panel) and $0.010 \pm 0.005$ (right panel); for \Pag, $0.009 \pm 0.003$ (left panel) and $0.008 \pm 0.004$ (right panel).
}
\vspace{0.3cm}
\label{fig:VoverR}
\end{figure*}

\begin{figure*}[t]
\includegraphics[width=\linewidth]{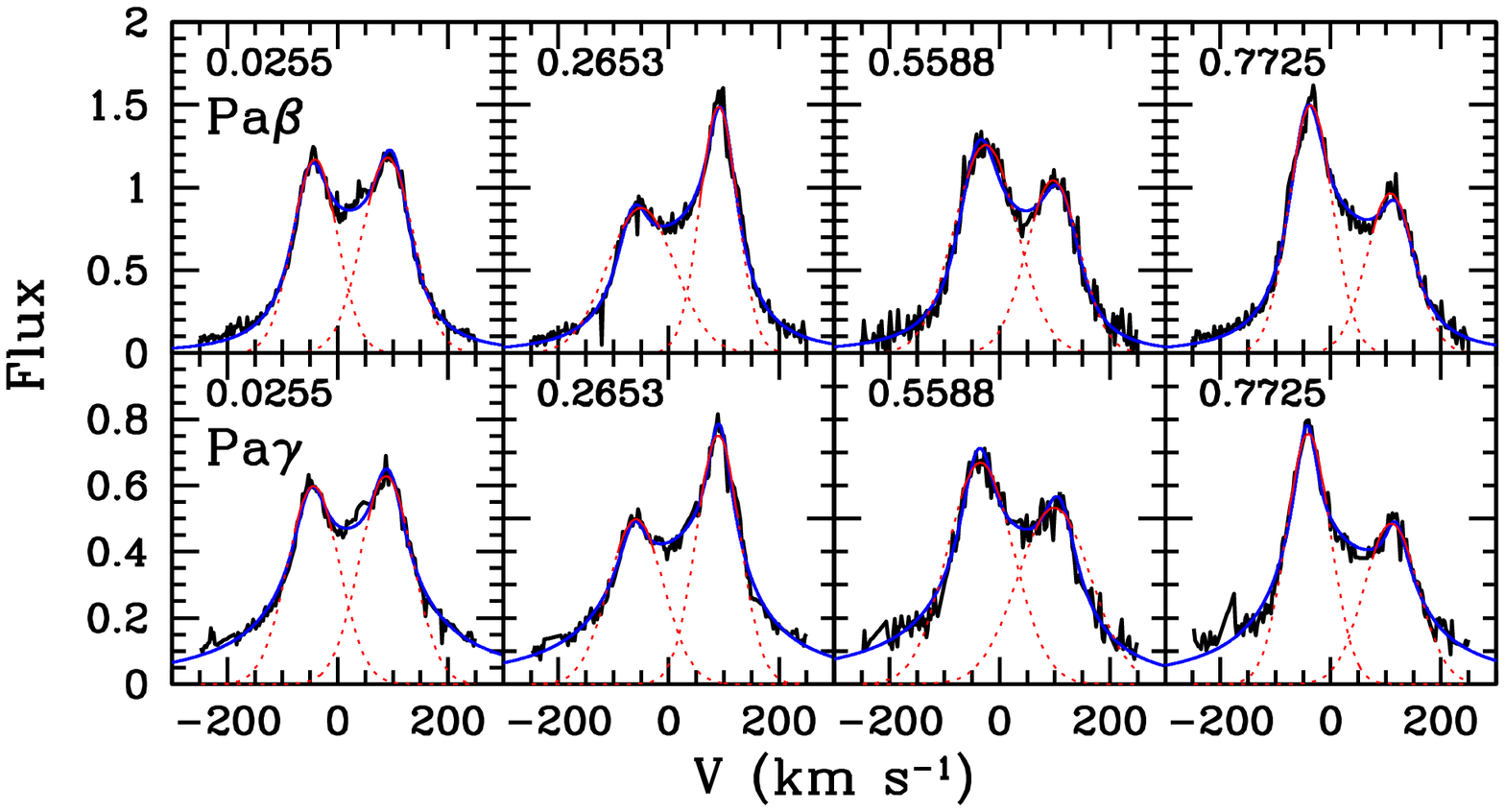}
\caption{
Example of spectral profiles that illustrate the V/R variations over the binary cycle, at four representative phases. The phases (labelled in each panel) are close to conjunction ($\phi \approx 0$, when the star is in front of the primary; and $\phi \approx 0.5$) and quadrature ($\phi \approx 0.25$ and $\phi \approx 0.75$). The top panels are \Pab\ profiles, the bottom panels show \Pag. In each panel, the black curve is the observed profile (after subtraction of the stellar absorption component), normalized in units of the stellar continuum near the line wavelength. The red curve around each peak is a Gaussian model, fitted only to the section of data corresponding to the interval of the solid red curve; the dotted part of the curve shows the full Gaussian function.  The blue curve is the best-fitting extended Smak profile (with free parameters for peak asymmetry).
}
\vspace{0.3cm}
\label{fig:ExampleFit}
\end{figure*}

The observed \Pab\ (12,821.6 \AA) and \Pag\ (10,941.2 \AA) profiles are double peaked at all orbital phases (Figure~\ref{fig:Pab}), and their stacks (Figure~\ref{fig:stacked}) are textbook examples of disk-like emission lines (``Smak profiles'': \citealt{smak81}; see also \citealt{horne86}). They are similar to the double peaked H$\alpha$ profiles observed from confirmed Galactic BHs in quiescence; for example, A0620$-$00 \citep{johnston89,orosz94}, GRS 1124$-$684 \citep{orosz94}, Swift J1357.2$-$0933 \citep{torres15}. The emission profiles of another Paschen line covered by our CARMENES spectra (Pa$\delta$ at 10,052.6 \AA) are also double peaked, but at lower signal-to-noise ratio than for \Pab\ and \Pag, and we will not use them here.

The simple shape of the Paschen line profiles is in stark contrast with the complex structure and multiple peaks of the \Ha\ line profiles L19. We interpret this difference as an effect of optical depth: we argue that \Ha\ is optically thick, and  includes contributions from multiple scattered photons as observed in nearly pole-on disks \citep{hanuschik96,hummel97}. In contrast, \Pab\ and \Pag\ are optically thin, and their profiles are not modified by scattering. Thus, the peaks are formed mainly by the photons emitted in the outer disk, and their velocity is a proxy for the projected rotational velocity at the outer edge of the disk \citep{paczynski77,smak81,horne86}; the wings are instead emitted from disk annuli at smaller radii. 

The extent of the wings provides a crucial test to determine whether the lines are emitted by a disk inside the Roche lobe of the primary object (hence, also tracing its orbital motion), or by a circumbinary disk (expected to be stationary). The wing velocity should not substantially exceed the projected rotational velocity of the inner edge of the disk (more exactly, the inner edge of the disk region in the right temperature range to emit a particular line). For a circumbinary disk, \cite{artymowicz94} computed an inner truncation radius $R_{\rm c} \approx 1.7 a$, where $a$ is the binary separation; slightly larger truncation radii, $R_{\rm c} \approx 2.0$--$2.3 a$, were derived in subsequent works \citep{holman99,macfadyen08,pichardo08}. 
In the LB-1 case, with a B-star radial velocity semi-amplitude of 52.6 \kms, the predicted inner-edge velocity of a circumbinary disk ranges from $\approx$40 km s$^{-1}$ for a mass ratio $\Mtwo/\Mone \approx 0.1$, to $\approx$70 km s$^{-1}$ for a mass ratio $\Mtwo/\Mone \approx 1$ \citep{blundell08}.

Our new spectral results show that the wings of \Pab\ and \Pag\ extend to more than $\pm$200 km s$^{-1}$ in the stacked profiles (Figure 2); indeed, the full-width at half maximum is $\approx$200 km s$^{-1}$ for both lines, already broader than possible for a circumbinary disk. Even in the individual spectra, it is clear that there is significant wing emission beyond $\pm$100 km s$^{-1}$ (Figure 1). Thus, we strongly confirm the argument presented by L19 (based on the width of the \Ha\ line) in favour of a Roche-lobe disk.

\subsection{Asymmetry and motion of the emission line peaks}

The phase sequence of the \Pab\ and \Pag\ profiles (Figure~\ref{fig:Pab}) shows that the relative flux of the two peaks varies with orbital phase (a phenomenon usually referred to as ``Violet/Red variations'' or ``V/R variations''). The red peak is higher than the blue peak during the half orbital phase in which the companion star is receding; the blue peak dominates when the star is approaching  (Figure~\ref{fig:VoverR}). Averaged over an orbital cycle, the emission from the two sides of the disks is consistent with being equal (Figure~\ref{fig:stacked}); the small residual asymmetry in the peak heights of our stacked profiles is simply due to a non-uniform phase coverage. 

A direct emission contribution to the asymmetry from the surface of the secondary star is ruled out because the radial velocity amplitude of the star (52.6 \kms) is significantly smaller than the velocity of the line peaks (with half separation $\sim$70 \kms). In addition, such a contribution would lead to line peaks always higher than those at phase $\phi\sim 0$ or 0.5, inconsistent with what is seen in the profiles (e.g., Fig.~\ref{fig:ExampleFit}).
There is also no evidence of an ``S-wave'' \citep{honeycutt87} moving across the two peaks, which is often seen in Cataclysmic Variables and is interpreted as the emission from the hot spot at the interface of accretion stream and disk. 
We suggest that the simplest explanation for the peak asymmetry and its variation with phase is that the gas in the disk is illuminated or heated by the B star, which increases the line emissivity on the disk side close to the star. We leave a detailed modeling of the effect of irradiation on the line profiles to future work. Here, we focus instead on the positions of the line peaks and their behaviour with orbital phase, regardless of the physical reasons for the alternate enhancement of the red and blue peak.

To measure the positions of the Paschen line peaks in a model-independent way, we used two methods. First, we fit each peak with a Gaussian. For this, we used a data interval of $\pm$35 \kms\ around the apparent position of each peak, at each orbital phase (Figure~\ref{fig:ExampleFit}). We take the center of each Gaussian fit as the peak position. The second method we used to determine the peak position is a fit to the analytic Smak profile \citep{smak81,horne86,johnston89} extended with two additional free parameters (amplitude and radial dependence) to describe emissivity asymmetry and thus the peak asymmetry. For this method, we considered the full velocity range from $-250$ km s$^{-1}$ to $+250$ km s$^{-1}$ in the observed velocity. 

Following \citet{paczynski77}, we then construct two quantities from the measured phase-dependent peak velocities. The first quantity, \Vd, is the mean of the red and blue peak velocity ($V_{\rm red}$ and $V_{\rm blue}$): $V_{\rm 1d} \equiv (V_{\rm red}+V_{\rm blue})/2$. The second one, \Vrot, is the half separation between the two peaks: $V_{\rm rot} \equiv (V_{\rm red}-V_{\rm blue})/2$. The quantity \Vd\ is a measure of the orbital motions of the disk around the center of mass of the system (a proxy for the primary orbital velocity), while \Vrot\ characterizes the rotation of the outer edge of the disk around the primary. 
We calculated \Vd\ and \Vrot\ for \Pab\ and \Pag, with two sets of values for each line (corresponding to the two fitting methods described above). 

\begin{figure*}[t]
\includegraphics[width=\linewidth]{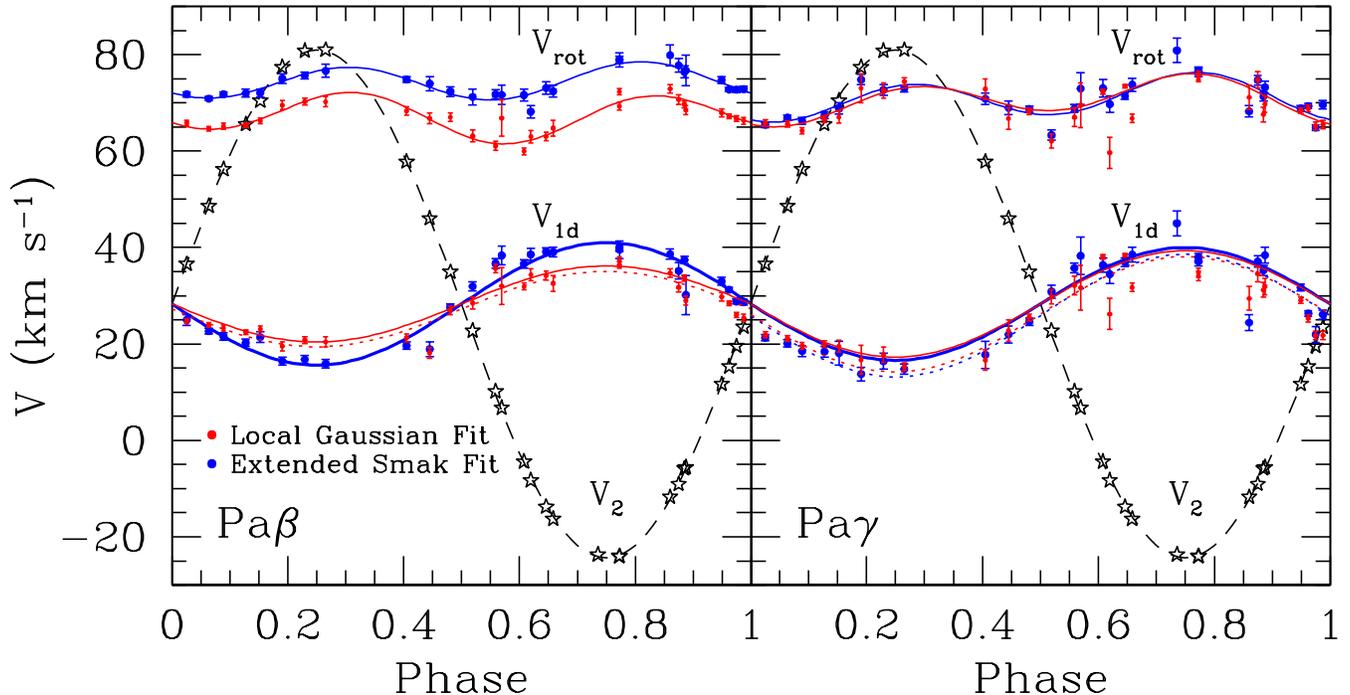}
\caption{
Measurements and model fitting of the projected radial velocities of the secondary star and primary, as a function of orbital phase. In both panels, the star symbols are the data points for the stellar motion, determined from template fitting to the \ion{He}{1} absorption lines; their best-fitting sinusoidal curve (dashed black line) has an amplitude $\Ktwo=52.62 \pm 0.15$ \kms\ and systemic velocity $V_{\rm sys} = 28.3 \pm 0.1$ \kms. Peak velocity measurements and modelling for the \Pab\ disk emission line are shown in the left panel, and for \Pag\ in the right panel. 
In each panel, the lower sets of red and blue data points, labelled \Vd, are the radial velocities of the midpoint between the emission peaks ($V_{\rm{1d}} = (V_{\rm {red}} + V_{\rm{blue}})/2$). The red data points correspond to peak positions obtained from local Gaussian fits, and the blue data points to the positions obtained from extended Smak profile fits. Each set of \Vd\ data points are fitted with two sinusoidal curves: the solid one has a baseline fixed to $V_{\rm sys}=28.3$ \kms, the dotted one leaves it as a free parameter. We take \Vd\ as a proxy for the orbital motion of the primary (more exactly, the observed semi-amplitude $\Kone \approx 0.85 K_{\rm{1d}}$ where $\Kone$ is the semi-amplitude of the  primary; see Section 4.1). The upper set of red and blue data points, labelled \Vrot, are the half-separations between the peaks as a function of orbital phase (red for Gaussian positions, blue for extended Smak profile positions). They are fitted with first-harmonic sinusoidal curves (i.e., with a period of half the orbital period), to account for the ellipsoidal distortions of the outer disk shape in the Roche lobe potential. \Vrot\ represents the projected rotational velocity of the disk edge. 
}
\label{fig:PabFit}
\vspace{0.3cm}
\end{figure*}

We find (Figure~\ref{fig:PabFit}) that for both \Pab\ and \Pag, regardless of fitting method, 
\Vd\ has a low-amplitude ($\sim$10 km s$^{-1}$) but clear sinusoidal behavior, in anti-phase with the B star, consistent with our interpretation as the orbital motion of the primary object (which must then be substantially more massive than the secondary star). The quantity \Vrot\ shows periodic modulations with half of the orbital period; this is consistent with the slightly non-circular shape of the gas streamlines (which determine the outer disk boundary) in the Roche lobe potential \citep{paczynski77}, as discussed in Section 4.

\section{Mass ratio from line modelling}

\subsection{Peak separation and orbital motion}

In the model of \citet{paczynski77}, the outer edge of the BH accretion disk is determined by the outermost closed streamline, under the influence of the binary potential in the rotating frame. The shape of the outer edge of the disk slightly deviates from a circle, and the amount of ellipticity depends on the binary mass ratio. This effect leads to a small difference between the amplitude of the fitted orbital motion \Vd\ and the true orbital motion of the primary; it also introduces phase-dependent variations in the peak separation \Vrot. 
To derive the mass ratio in the framework of this model, we fit \Vd\ with a sinusoidal curve, $\propto \sin (2\pi \phi$), and \Vrot\ with linear combinations of $\sin (4\pi \phi$) and $\cos (4\pi \phi$), where $\phi\in [0, 1]$ is the binary phase. We applied this model first to \Pab\ and then to \Pag, for both sets of peak values (from Gaussian and extended Smak fits).

For \Pab, using the extended Smak fitting function, the mean peak position \Vd\ has a semi-amplitude $K_{\rm 1d}=12.7 \pm 0.3$ \kms, with the baseline fixed to the systemic velocity of 28.3 \kms\ determined from the B-star radial velocity. In the initial run of our model, the phase of the sinusoidal curve for the disk motion was left free: we obtained a best-fitting phase value of $0.006\pm 0.004$, that is almost perfectly in anti-phase with the motion of the companion star. Based on this close match, we then fixed the phase angle to $0$ for our subsequent modelling, to reduce the number of free parameters. The mean peak half-separation is $\langle V_{\rm rot}\rangle=74.4 \pm 0.2$ \kms. The value of $K_{\rm 1d}/\langle V_{\rm rot}\rangle\approx 0.171$ corresponds to a mass ratio $\Mone/\Mtwo = 4.0 \pm 0.1$ \citep[][Table 2]{paczynski77}. The same model also provides a correction factor to convert \Kd\ to the semi-amplitude $\Kone$ of the primary radial velocity: we infer $\Kone = 11.2 \pm 0.3$ \kms. If we compare this value with the radial velocity semi-amplitude $\Ktwo = 52.6 \pm 0.2$ \kms\ of the star, we obtain an alternative estimate of the mass ratio $\Mone/\Mtwo = 4.7 \pm 0.2$. 

Also from the extended Smak profile fitting of \Pab, we find a semi-amplitude $\Delta V_{\rm rot}\approx 3.6$ \kms, corresponding to $\Delta V_{\rm rot}/\langle V_{\rm rot}\rangle \approx 0.048$. According to Table 2 in \citet{paczynski77} with the correction factor 2 mentioned in the note added in proof, this would imply an high mass ratio of about 12.4. We suggest that this discrepancy is explained if viscosity shapes the outer edge of the disk to be more circular than in the pure streamline model of \citet{paczynski77}, leading to a smaller amplitude in the orbital variation of \Vrot. We also find a phase shift of $\sim$41$^{\circ}$ in the variation, compared to the model in \citet{paczynski77}. Therefore, we do not use the orbital variations of \Vrot\ to infer the mass ratio. A more circular outer disk would reduce the correction factor between the observed $K_{\rm 1d} \approx 12.7$ \kms\ and the true orbital semi-amplitude of the primary; in the limiting case that $K_{\rm 1d} = \Kone$, the mass ratio $\Mone/\Mtwo = \Ktwo/\Kone \approx 4.1$, still in the same ballpark as the values inferred above.

We repeated the same analysis for the \Vd\ and \Vrot\ from Gaussian fitting of the \Pab\ peaks, and for \Pag\ with both fitting methods. In total, we obtain eight alternative values of mass ratio $\Mone/\Mtwo$ (Table 2). Our full range of estimates spans $\approx$3.9--8.4 (including their 68\% error ranges), with a mean value of $\approx$5.3 and a 1-$\sigma$ dispersion of $\approx$1.4. 
If we remove the largest and the smallest values, we obtain a range of $\approx$4.2--6.5 with a mean value of $\approx$5.1 and a 1-$\sigma$ dispersion of $\approx$0.8.

\input{tab_v.tex}

\subsection{Outer radius of the disk}

For the observed mass ratio $\Mone/\Mtwo \approx 5$, the tidal truncation radius of the disk \citep{paczynski77,warner95} is $r_{\rm d} \approx 0.60 \, a \Mone/(\Mone +  \Mtwo) \approx 0.50 \,a$, where $a$ is the binary separation. An alternative definition for the outer edge of the disk is the 3:2 resonance radius \citep{whitehurst91}, which is $\approx$0.45\,$a$ for a mass ratio of $\approx$5. 

To determine whether the observed disk parameters are consistent with those predictions, first we used the simplifying approximation of a circular outer disk (radius $r_{\rm d}$) in Keplerian rotation around the primary (i.e., rotating with a speed $v = \sqrt{GM_1/r_{\rm d}}$). The inferred $\langle V_{\rm rot}\rangle$ and mass ratio, in combination with the observed radial velocity semi-amplitude of the secondary star ($\Ktwo$), enable us to determine $r_d/a$. Somewhat surprisingly, we obtain a large value $r_{\rm d} \approx 0.6\,a$, even larger than the expected volume-averaged radius of the Roche lobe ($r_{\rm L} \approx 0.52 \, a$: \citealt{eggleton83}). However, the circular Keplerian approximation is too simplistic for an accurate estimate of the outer disk radius in the Roche lobe geometry. When the proper gravitational potential of the binary is taken into account \citep{huang1967}, the rotational speed in the outer rings of a large disk is lower than the Keplerian speed of rotation around an isolated body of mass $M_1$; in other words, the Keplerian approximation over-estimates the outer disk radius. We used a fitting formula to the values in Table 2 of \citet{huang1967} to derive a more accurate disk size. With this method, we find an outer radius $r_{\rm d} \approx 0.36 \, a$ (Table~\ref{tab:velocity}). This is smaller than the tidal truncation radius calculated by \citet{paczynski77}. As the orbits in the disk are expected to be more circular at smaller radii, this result suggests that we may have over-corrected $K_{\rm 1d}$ to obtain $\Kone$ and thus we may have over-estimated the mass ratio $\Mone/\Mtwo$. However, this effect is small (only a few percent), and even in the implausibly extreme case of $\Kone=K_{\rm 1d}$, the estimated mass ratio is still well within the range of $\approx$4--7 we derived before (Section 4.1).

\subsection{Line peak versus line wing modelling}

In this work, we have traced the orbital motion and calculated the mass ratio from the sinusoidal shift of the Paschen emission line peaks, with two alternative methods for the identification of the peaks and two alternative conversions between peak position and mass ratio. Systematic uncertainties include the amount of smearing of the peaks (which tends to reduce the apparent peak separation: \citealt{marsh94}), the tidal distortion and possible eccentricity of the outer disk \citep{smak97}, and perhaps most importantly, the mean radius of the outer disk edge. The last factor could be either the largest non-intersecting streamline, or the 3:2 resonance radius, or perhaps, as we suggested here (Section 4.2), the 2:1 resonance radius. In principle, emission line profiles could also be affected by S-wave contributions of accretion stream and hot spot \citep{wade85,kaitchuck94}, or emission enhancements from spiral density waves in the disk \citep{steeghs99}; however, we do not see any evidence of such terms in our observed Paschen line profiles. What we do see, instead, is a periodic enhancement of the flux from the red or blue side of the line profile, in phase with the position of the secondary star. We suggested that this effect is caused by the asymmetric irradiation of the gas in the outer disk by the secondary star.

Instead of tracing the orbital motion of the disk from its emission line peaks, we can do it by fitting the wing position. The wing fitting was the method used by L19, on the H$\alpha$ line profile, which does not show clear double-peaks as in Paschen lines. In principal, the wings are emitted from smaller disk radii, thus more sensitive to the orbital motion of the primary and less affected by the tidal distortion of the outer disk or by hypothetical S-wave emission. However, wing fitting has its own downsides compared with peak fitting. First, 
the emission wing position is more sensitive to the accurate subtraction of the pressure-broadened absorption wings of the stellar line (which moves in antiphase with the disk line). The exact profile of the stellar absorption line depends on the temperature and surface gravity adopted for the stellar template. The second problem is that wing profiles are more easily affected by stellar and disk winds. Third, for LB-1, the alternate enhancement of blue and red peaks in phase with the stellar motion, from the stellar irradiation effect, causes a corresponding displacement of the wing positions measured at a fixed fraction of peak flux; if not properly accounted for, this effect leads to cancellation or severe under-estimate of the true orbital motion of the primary.
Both \cite{Abdul2019} and \cite{elbadry20} showed that the stellar absorption term may cause an apparent motion of $\sim$6 km/s mimicking that measured by L19, then the contributions of stellar winds and irradiation effects, not considered in neither \cite{Abdul2019} nor \cite{elbadry20}, may cause an opposite apparent motion of $\sim$6--10 km s$^{-1}$ in the line wings based on the primary motion from the Paschen line peaks. 

In contrast, flux asymmetry does not substantially affect the peak position, which is tied to the location of the outer disk edge. Variations in relatively broad spectral features, such as the stellar absorption profile and possible wind emission features, should also have little effect on the velocity amplitude \Vd\ of the mean peak position\footnote{Instead, the observed peak separation \Vrot\ may be substantially reduced by an optically thick disk wind \citep{murray97}.}.  This is why in this work we focused on peak fitting, considering its results a more reliable indicator of the binary mass ratio.  We leave a wing-fitting study of the same set of Paschen lines in combination with the H$\alpha$ line to a follow-up work currently in preparation.

\section{Discussion}
\label{sec:discussion}

Our measurement of the mass ratio must be combined with the mass estimate for the secondary star, based on its temperature, luminosity, and surface gravity. Below we briefly discuss the implications of this work for different scenarios proposed for LB-1, but leave definitive tests to future work.

\subsection{a B star plus a BH}

Two types of solutions, corresponding to two classes of evolutionary tracks, appear to be consistent with the observed B3-type colours and spectra. The first class of solutions represents stars that have already moved off the main sequence, but not yet reached the giant stage. Such stars are moving from higher to lower temperatures in the Hertzsprung-Russell diagram. The second class of solutions represents stars that have been partially stripped of their hydrogen envelopes during close binary evolution.

For the first class, L19 proposed a subgiant mass $\Mtwo = 8.2 \pm 0.9\, M_{\odot}$ based on the {\sc{tlusty}} model \citep{hubeny95} fitting to Keck/HIRES spectra (resulting in $T_{\rm {eff}} = 18,100 \pm 820$ K, $\log g = 3.43 \pm 0.15$) and the PARSEC evolutionary tracks. Several works re-estimated the atmospheric parameters using new high-resolution spectral observations or archive data. For example, \cite{Simon2020} derived $T_{\rm eff} = 14,000 \pm 500$ K and $\log g = 3.5 \pm 0.15$, using the non-LTE stellar atmosphere code {\sc{fastwind}}. These spectroscopic parameters corresponds to 
$\Mtwo = 5.2^{+0.3}_{-0.6} M_{\odot}$ by comparison with the evolutionary tracks of \cite{ekstrom12} and \cite{brott11}. 
Using the Grid Search in Stellar Parameters fitting suite \citep{tkachenko15}, \cite{Abdul2019} modelled the spectrum with an effective temperature $T_{\rm eff} = 13,500 \pm 700$ K and a surface gravity $\log g = 3.3 \pm 0.3$; this corresponds to a mass $\Mtwo = 4.2^{+0.8}_{-0.7} M_{\odot}$, according to the evolutionary tracks from \cite{brott11}.

To summarize, several groups have obtained a mass range of 3--6 $M_\odot$ based on spectroscopy, lower than the L19 results.
Note that the mass estimate depends on the models adopted. 
For example, if PARSEC evolutionary tracks are adopted instead, the ($T_{\rm eff}$, $\log g$) values in \cite{Abdul2019} would correspond to $5.6^{+1.3}_{-1.2} M_\odot$ instead of $4.2^{+0.8}_{-0.7} M_{\odot}$, as similarly shown in Fig.~2 of \cite{Simon2020}.
Nevertheless, if the secondary star is indeed a B star with a conservative mass estimate of 3--6$M_\odot$, then the mass of the primary is about 13--36$M_\odot$ (12--50$M_\odot$) for our new mass ratio of $5.1\pm0.8$ (3.9--8.4).

\subsection{A stripped helium star plus a BH}
 
The second class of solutions are moving from the redder to the bluer part of the Hertzsprung-Russell diagram on their way to the subdwarf stage \citep{gotberg18}; thus, they cross the main-sequence to subgiant tracks. The mass of such stars is much lower than the mass of a main sequence or subgiant with similar temperature and surface gravity. Spectral modelling of LB-1 by \cite{Irrgang2020}, by \cite{Simon2020} and by \cite{shenar20} highlighted the super-solar He abundance and the presence of CNO-processed material (enhanced in N and depleted in C and O), normally not present near the surface of main sequence or subgiant stars. This suggests that the star has been partially stripped of its hydrogen envelopes during close binary evolution.

Using a modified version of the Binary-Star Evolution code \citep{hurley02},  \cite{yungelson20} proposed that the secondary star (with current mass $0.5 \lesssim M_2 \lesssim 1.7 M_{\odot}$) had an initial mass $\sim$4--9$M_{\odot}$ but is now in its He-shell burning phase, after a phase of case-B Roche lobe overflow in which it lost most of its envelope to the primary. \cite{yungelson20} estimated an age of the system of $\approx$80 Myr, and a duration of the He-shell burning phase of $\approx$7 Myr.  Similarly, using the Binary Population and Spectral Synthesis ({\sc {BPASS}}) V2.2 models \citep{eldridge17,stanway18}, \cite{eldridge20} found acceptable solutions for the secondary star at two different stages (separated by $\approx$25 Myr) of its evolution towards the subdwarf (and eventually white dwarf) stage. An analogous suggestion about the evolutionary track of the He-burning star was put forward by \cite{Irrgang2020}, based on the binary models of \cite{gotberg18}. Using the PoWR model, we derived an effective temperature $T_{\rm eff} = 13,500$ K, a surface gravity $\log g = 2.8$, and a luminosity $L=10^{3.1} L_{\odot}$, with $d = 2.14$ kpc and $E(B-V) = 0.6$ mag. This corresponds to a companion mass $\Mtwo \approx 1 M_{\odot}$.

In summary, although the secondary star may have started its life on the main sequence as a $\sim$4--9$M_{\odot}$ B-type star, abundance ratios support the view that its mass is now only $\sim$1--2 $M_{\odot}$. 
If the secondary is indeed such a stripped helium star, the mass of the primary would be 4--12$M_\odot$ (4--17$M_\odot$) for 
our new mass ratio of $5.1\pm0.8$ (3.9--8.4).

\subsection{A stripped helium star plus a Be star}

\cite{shenar20} proposed that the optical continuum in LB-1 is the superposition of an almost non-rotating stripped star and a fast-rotating B3e main sequence star. Note that the definitions of primary and secondary in \citet{shenar20} are opposite to ours. To be consistent within this paper, we adopt our definitions here to discuss their results, with the fast-rotating B3e star as the primary and the almost non-rotating stripped star as the secondary.  Their subsequent decomposition finds a B3Ve star with strong H$\alpha$ emission line that moves at $K_1 = 11\pm1 $ \kms (interestingly close to our estimate in this paper), and a stripped helium star with narrow absorption lines that move at 53 \kms as reported by L19. By calibrating the mass of the B3Ve primary to $M_1 = 7\pm2M_\odot$, \cite{shenar20} derived an orbital mass for the stripped secondary of $M_2 = 1.5\pm0.4M_\odot$.
 
In this scenario, the primary is a fast rotator (which explains its decretion disk) because it has been spun up by mass and angular momentum transfer from the stripped secondary.  The archetype of this class of systems is $\varphi$ Per \citep{gies93,gies98,hummel01,schootemeijer18}. Other systems in this class (five confirmed plus a dozen strong candidates) are discussed in \cite{schootemeijer18} and \cite{wang18}. When reliable measurements for the stripped secondary are available, such systems typically show temperatures $\gtrsim$25,000 K (hence the alternative name Be $+$ sdO systems, the stripped star having an O-type spectrum). However, it was suggested \citep{schootemeijer18,wang18} that the known Be $+$ sdO systems are only the high-luminosity end of the distribution, and a much larger (so far undetected) population with fainter, cooler secondary stars should also exist. LB-1 could be an example of this population, in which the secondary is only partially stripped and has an effective temperature of $\approx$13,000 K.

Such a scenario, however, relies on  the decomposition of the optical spectra, which is a reverse problem that can hardly have a unique solution. For example,  their decomposition led to double-peaked Balmer emission lines (as seen in their Figure~1) from the cold stripped helium star, which is hard to explain for a cold secondary that is far from filling its Roche lobe. On the other hand, this could be residuals given the low S/N of their spectra data and the neglect of irradiation effect in the spectral disentangling. We will combine PoWR modeling and our high S/N CARMENES spectra data to verify such a decomposition in a future work. Solid evidence for this intriguing scenario has to wait for Gaia astrometric data as discussed later.

\subsection{A triple star system}

\cite{rivinius20} studied in depth HR6819 that appears as a bright early Be star but can be decomposed with an additional B3III component. Their spectroscopic time series taken in 2004 revealed the narrow absorption lines in its B3III component, which exhibit a period of 40 days in a circular orbit, and also broad H$\alpha$ emission lines without apparent motion, stunningly similar to the observations of LB-1. \cite{rivinius20} interpreted HR6819 as a triple system, where the line emission comes from the disk around a distant third body in the system: a Be star (\citealt{rivinius13} for a review) with a decretion disk, while the B3III motion comes from an inner binary with a BH above $4M_\odot$. 

\cite{rivinius20} suggested that LB-1 is also such a triple star system, and the mass of the primary in LB-1 can then be reduced from $\sim70M_\odot$ to a level more typical of Galatic stellar remnant BHs.  In this scenario, the orbital motion of the third body would be too slow and its period too long ($>$1 year) to show up in the spectroscopic observations to-date. However, as our CARMENES spectra clearly show, the emission lines in LB-1 are actually moving in anti-phase with the absorption lines, thus cannot come from a distant third Be star as proposed for HR6819. In any case, short-cadence, high-precision radial velocity monitoring of such systems helps constrain on their nature \citep[e.g.,][]{hayashi20}.

\section{conclusions}

Mass ratio estimates in LB-1 based on studies of the H$\alpha$ emission line are fraught with difficulties, because of the distorted line profile, probably affected by electron scattering and by other emission components in addition to a standard disk. To reduce those difficulties, we analyzed the \Pab\ and \Pag\ emission lines, observed with the CARMENES spectrograph at Calar Alto Observatory. After proper subtraction of the underlying stellar absorption, the phase-averaged Paschen lines show a textbook double-peaked Smak profile. The red and blue side of the line are alternately enhanced during the 78.9-d orbital cycle, in phase with the secondary star; we attributed this effect to stellar irradiation of the gas on either side of the disk. We have measured the peak position and modelled their phase dependence, to determine the size of the disk and the mass ratio of the system. 

We found four observational pieces of evidence that point to a disk located inside the Roche lobe of the primary, rather than a circumbinary disk as suggested by \cite{elbadry20}, or a disk around a hierarchical third body as suggested by \cite{rivinius20}. First, there is a clear sinusoidal variability of the mid-position between the peaks, with the same binary period and in antiphase with the motion of the companion star.
Second, the peak half-separation ($\sim$70 km s$^{-1}$), proxy for the projected rotational velocity at the outer disk edge, is larger than the radial velocity amplitude of the companion star ($\approx$53 km s$^{-1}$). 
Third, the peak-to-peak separation has an ellipsoidal component with a period of half the binary period, which is expected for tidal deformation of the outer disk inside the primary Roche lobe. 
Fourth, the line wings in the emission line profile extend at least to a velocity of $\pm$250 km s$^{-1}$ either side of the systemic velocity, much in excess of what is expected from a circumbinary disk.

From the sinusoidal motion of the line peaks, compared with the amplitude of the stellar motion, we have inferred a mass ratio of $\approx$3.9--8.4 (including their 68\% error ranges), with a mean and 1-$\sigma$ error of  $\Mone/\Mtwo = 5.3 \pm 1.4$, or $\Mone/\Mtwo = 5.1 \pm 0.8$ if we reject the highest and lowest estimate from our analysis. This value is lower than the mass ratio of $\approx$8 claimed in L19. The three main differences between our results and those of L19 are that we are using Paschen lines with a cleaner disk profile, we have properly accounted for the stellar absorption component, and we have fitted the peak rather than the wing positions. The (average) outer edge of the disk is located at a radius $r_{\rm d} \approx 0.36 a$ (derived using a proper Roche-lobe gravitational potential rather than a simple circular Keplerian approximation for the outer disk edge).

Our measurements of the emission line motion and thus the inferred mass ratio ($5.1\pm 0.8$) can be combined with the recently proposed mass and spectral type of the secondary star.
If the secondary star is indeed a B star with a conservative mass estimate of 3--6$M_\odot$, then the mass of the primary is about 13--36$M_\odot$.
If the secondary is indeed a stripped helium star of 1--2$M_\odot$, the mass of the primary would be 4--12$M_\odot$. Our disk line analysis cannot distinguish between a BH primary with an accretion disk, and a Be primary of similar mass, with a decretion disk as proposed by \cite{shenar20}. On the other hand, the clear sinusoidal motion of Paschen emission lines argues strongly that LB-1 is not a triple system as proposed by \cite{rivinius20}.

It is intrinsically difficult to determine masses from spectroscopy alone in massive binaries, as in the case of LB-1, let alone disentangle the possible contribution of two stars. Different results are obtained from different datasets used for spectral modelling, and from different stellar atmosphere and evolution models used by various groups. An independent estimate of the primary mass will soon be obtained with reduced uncertainties thanks to the astrometric data from the {\it Gaia} mission \citep{Gaia2016}. As pointed out in L19, {\it Gaia} transit data in the next Data Release will reach an astrometric error as small as 0.1 milli-arcsecond, enough to resolve the binary wobble of LB-1. The full orbit and the parallax can be solved simultaneously from the combination of radial velocity measurements and {\it Gaia} transit data. This will give the total binary mass and the mass ratio, simultaneously. 

Moreover, for the same total mass and mass ratio, {\it Gaia}'s astrometric precision may also distinguish between the BH and Be star scenarios. For example, let us assume $M_2 = 1.5 M_{\odot}$, $M_1/M_2 = 5$. In a BH system, the photometric center coincides with the secondary star (the disk having a negligible contribution to the optical continuum) and rotates with a semi-amplitude of $\approx$0.62 AU, corresponding to $\approx$280 $\mu$as at the distance of 2.2 kpc. In a Be system with the same masses, the photometric center is approximately half way between the two components \citep{shenar20}, and hence moves with a semi-amplitude of only $\approx$0.25 AU $\approx$110 $\mu$as.  

In conclusions, the original motivation of our LAMOST survey was to search for a hidden population of binary systems with X-ray-quiescent BHs. With LB-1, we have either found an example of such population (although with a much more reasonable mass than originally inferred in L19), or we have stumbled upon an example of another largely hidden population of Galactic binaries (Be star plus stripped He-burning companion). Both types of systems can be discovered in bulk from long-term spectroscopic monitoring of a large number of stars, as we are doing with the LAMOST survey.

\acknowledgments

We thank Drs. Sergio Simon-Diaz, Stephen Justham, Helen Johnston, Christian Motch, and many others for constructive suggestions and comments for this work. 
The Calar Alto Observatory is jointly run by the Institute of Astrophysics of Andalusia (IAA-CSIC) and the Council of Andalusia. This work was supported by the National Science Foundation of China (NSFC) under grant numbers 11988101/11933004 (J.L.), 11690024 (Y. L.) and 11603035 (S.W.). RS thanks the University of Sydney and the South African Astronomical Observatory for hospitality during part of this work.

\bibliography{main_arxiv.bbl}{}
\bibliographystyle{aasjournal}

\end{document}

%% file: tab_v.tex
\begin{deluxetable*}{ccrrrcccc}
\tabletypesize{\scriptsize}
\tablewidth{0pt}
\tablenum{2}
\tablecaption{Kinematics and mass ratios inferred from the disk emission lines
\label{tab:velocity}}
\tablehead{
\colhead{Model} & \colhead{Line}& \colhead{$K_{\rm 1d}$} & \colhead{$\langle V_{\rm rot}\rangle$} &
\colhead{$\Kone$} & \colhead{$\Mone/\Mtwo|_P$} & \colhead{$\Mone/\Mtwo|_V$} & $r_d/a|_P$ & $r_d/a|_V$\\
\colhead{}      & \colhead{}    & \colhead{($\rm km~s^{-1}$)}& \colhead{($\rm km~s^{-1}$)} & 
\colhead{($\rm km~s^{-1}$)} & \colhead{}               & \colhead{}                    & \colhead{}  & \colhead{}
}
\colnumbers
\startdata
{Extended Smak Fit}  & Pa$\beta$  & $12.7\pm 0.3$ & $74.4\pm 0.2$ & $11.2\pm 0.3$ & $4.04\pm 0.11$ & $4.70\pm 0.13$ & $0.347\pm 0.001$ & $0.345\pm 0.001$\\
{                 }  & Pa$\gamma$ & $11.7\pm 0.3$ & $70.9\pm 0.2$ & $10.2\pm 0.4$ & $4.23\pm 0.16$ & $5.15\pm 0.19$ & $0.362\pm 0.001$ & $0.360\pm 0.001$\\
\hline
{Local Gaussian Fit} & Pa$\beta$  & $ 7.9\pm 0.2$ & $67.4\pm 0.2$ & $ 6.5\pm 0.2$ & $6.51\pm 0.23$ & $8.11\pm 0.27$ & $0.375\pm 0.001$ & $0.374\pm 0.001$ \\
{                 }  & Pa$\gamma$ & $11.0\pm 0.3$ & $70.7\pm 0.2$ & $ 9.6\pm 0.3$ & $4.53\pm 0.14$ & $5.50\pm 0.15$ & $0.363\pm 0.001$ & $0.361\pm 0.001$\\
\enddata
\tablecomments{
The projected radial velocity amplitude of the primary, $\Kone$, in column (5), is defined as the observed disk line velocity amplitude $K_{\rm 1d}$ in column (3), corrected by a factor given in the model of \cite{paczynski77}. The mass ratio in column (6) is obtained from the observed values of $K_{\rm 1d}/\langle V_{\rm rot}\rangle$, compared with the values tabulated by \cite{paczynski77}. The mass ratio in column (7) is inferred from the ratio of the secondary to primary velocity amplitudes, $\Ktwo/\Kone$. Columns (8) and (9) are the outer disk size in units of binary separation, estimated based on the ratio of the disk outer edge velocity $\langle V_{\rm rot}\rangle$ and the secondary velocity $\Ktwo$, where the radial dependent disk velocity is assumed to follow the sub-Keplerian solution in \citet{huang1967}, with mass ratios from column (6) and (7), respectively.
}
\end{deluxetable*}